\documentclass[12pt,preprint]{aastex}
\begin{document}

\title{  The History of the Mysterious Eclipses of KH~15D:\\
         Asiago Observatory, 1967--1982$~^1$ }

\author{ John Asher Johnson\altaffilmark{2},
	 Joshua N.\ Winn\altaffilmark{3,4} }

\email{  johnjohn@astron.berkeley.edu;
         jwinn@cfa.harvard.edu }

\altaffiltext{1}{Based on data from the digitized Italian photographic
archives, produced under contract MIUR/COFIN 2002 to C.\ Barbieri,
Department of Astronomy, University of Padova.}

\altaffiltext{2}{Department of Astronomy, University of California,
Mail Code 3411, Berkeley, CA 94720}

\altaffiltext{3}{Harvard-Smithsonian Center for Astrophysics, 60
Garden St., Cambridge, MA 02138}

\altaffiltext{4}{National Science Foundation Astronomy \& Astrophysics
Postdoctoral Fellow}

\begin{abstract}
We are gathering archival observations to determine the photometric
history of the unique and unexplained eclipses of the
pre--main-sequence star KH~15D. Here we present a light curve from
1967--1982, based on photographic plates from Asiago
Observatory. During this time, the system alternated periodically
between bright and faint states, as observed today. However, the
bright state was 0.9~mag brighter than the modern value, and the
fractional variation between bright and faint states ($\Delta I =
0.7$~mag) was smaller than observed today (3.5~mag). A possible
explanation for these findings is that the system contains a second
star that was previously blended with the eclipsing star, but is now
completely obscured.
\end{abstract}

\section{Introduction}

Pre--main-sequence stars are well known for their photometric
variability, both erratic and periodic
\citep[e.g.][]{hhg+94}. However, the photometric behavior of KH~15D is
unusual even by the standards of pre--main-sequence stars. The star is
a 2--10 Myr-old T~Tauri star of spectral type K7, located in the open
cluster NGC~2264 \citep{hhs+01}. As discovered by \citet{kh98}, the
star's brightness decreases by 3.5 magnitudes every 48.35 days, and
these deep brightness excursions currently last more than 20 days. The
extreme depth and long duration of the eclipses rule out occultation
by a companion star, or modulation by star spots, as possible
explanations. The strict periodicity seems incompatible with a
mechanism based on unstable accretion. During brightness minima the
fractional polarization rises, suggesting that the entire face of the
star is occulted by circumstellar material and that only scattered
light is received at those times \citep{abw+03}. It is hoped that
continued observations will reveal the geometry and structure of the
material and provide unique information about circumstellar processes,
or even planet-forming processes.

The brightness variations are not accompanied by significant color
variations \citep{hhv+02}, implicating large particles or macroscopic
bodies as the occultors.  The ingress and egress light curves can be
understood qualitatively as the result of a sharp occulting edge
moving across the face of the star with a velocity vector nearly
parallel to the edge. A peculiar phenomenon observed at mid-eclipse is
the ``central re-brightening'': a phase lasting a few days during
which the star re-brightens. In the past, the central re-brightening
returned the system brightness to the out-of-eclipse level. In one
early observation, the star {\it exceeded}\, its uneclipsed brightness
by 0.5~magnitude \citep{hhs+01}.

Ongoing photometric monitoring by \citet{hhs+01} and \citet{hhv+02} has
shown that the light curve of KH~15D evolves with time. The duration
of the eclipses has been growing steadily since 1996, at a rate of
$\approx$1~day~year$^{-1}$.  The central re-brightenings have been
declining in strength, and are now much less pronounced than they used
to be. Given these secular changes, it should be illuminating to
assemble a historical light curve of KH~15D from archival photographic
observations of NGC~2264. Fortunately, it is realistic to expect that
suitable archival observations exist. The star is presently bright
enough ($V=16$) to have been detected on photographic plates with
modest-sized telescopes. It resides in a young and relatively
unobscured cluster full of variable stars that has attracted
scientific interest for many decades.

Winn, Garnavich, Stanek, \& Sasselov (2003) performed the first
archival study of KH~15D, using plates from the collection of the
Harvard College Observatory. Although the star was too faint and too
blended with a nearby bright star for accurate photometry,
\citet{wgs+03} established that between 1913 and 1955, the star was
rarely (if ever) fainter by more than one magnitude than its present
uneclipsed state.  They expressed this result as an upper limit of
20\% on the duty cycle of 1~mag eclipses.

This showed that the deep eclipses are a recent phenomenon, and
motivated the search for more photographs taken after 1950 in order to
identify the onset of the eclipses. We have been gathering additional
plates from observatories around the world in order to fill in the
photometric history of KH~15D. This paper presents new results from
our archival exploration, based on a time series of high-quality
photographic plates from the Asiago Observatory, taken between 1967
and 1982. In \S~\ref{data}, we describe the plates and the
digitization process. We determined the magnitude of KH~15D on each
plate with the procedure described in \S~\ref{photometry}. The Asiago
light curve is presented in \S~\ref{results} and compared to the
modern light curve. We conclude in \S~\ref{discussion} by summarizing
the results and speculating on the reason for the striking evolution
of the system between 1967 and today.

\section{Selection and digitization of the plates}
\label{data}

The Astrophysical Observatory of Asiago\footnote{{\tt
http://www.pd.astro.it/asiago/}}, in northern Italy, houses a
collection of nearly 80,000 photographic plates \citep{bor03}. Among
them are more than 300 images of NGC~2264, but most of the exposures
were too shallow to expect KH~15D to be detectable. In some other
cases, the exposure is sufficiently deep, but the position of KH~15D
is excessively contaminated by the extended halo of scattered light
from a nearby bright B star, HD~47887. This was also the main problem
with the Harvard plates (Winn et al.\ 2003).

All the plates described in this paper were obtained with 92/67~cm
Schmidt telescope at the Cima Ekar station of Asiago Observatory,
between 1967 and 1982. We found 48 plates with reasonably long,
red-sensitive exposures that were ideally suited for our study. These
were exposed with a I-N emulsion and RG5 filter. The I-N emulsion was
designed for improved sensitivity redward of typical astronomical
emulsions, to a cutoff near 9200\AA~\citep{sw62}. The long-pass RG5
filter is identical to the modern RG665 Schott filter and blocks
wavelengths shorter than 6400\AA. Together, the I-N/RG5 combination
has a transmission curve similar to the $\approx$7000--9000\AA~band
pass of the Cousins $I$ band.

The plates with bluer sensitivity were not as well-suited for accurate
photometry, but we wanted to get at least some color information, so
we also searched for a few high-quality examples.  We selected 3
plates exposed with a 103a-E emulsion and RG1 filter, which together
have a spectral response similar to the Johnson $R$ band \citep{mm00}.
We also selected one plate exposed with a 103a-O emulsion and GG5
filter, giving a spectral response similar to the Johnson $B$
band. The contents of our Asiago sample can be summarized as 48
$I$-like plates, 3 $R$-like plates, and one $B$-like plate. These fall
into three groups in time, with 41 plates taken from 1967 to 1970, 6
plates from 1973 to 1976, and 5 plates from 1979 to 1982.

\begin{figure}[p]
\begin{center}
\includegraphics[width=1\textwidth]{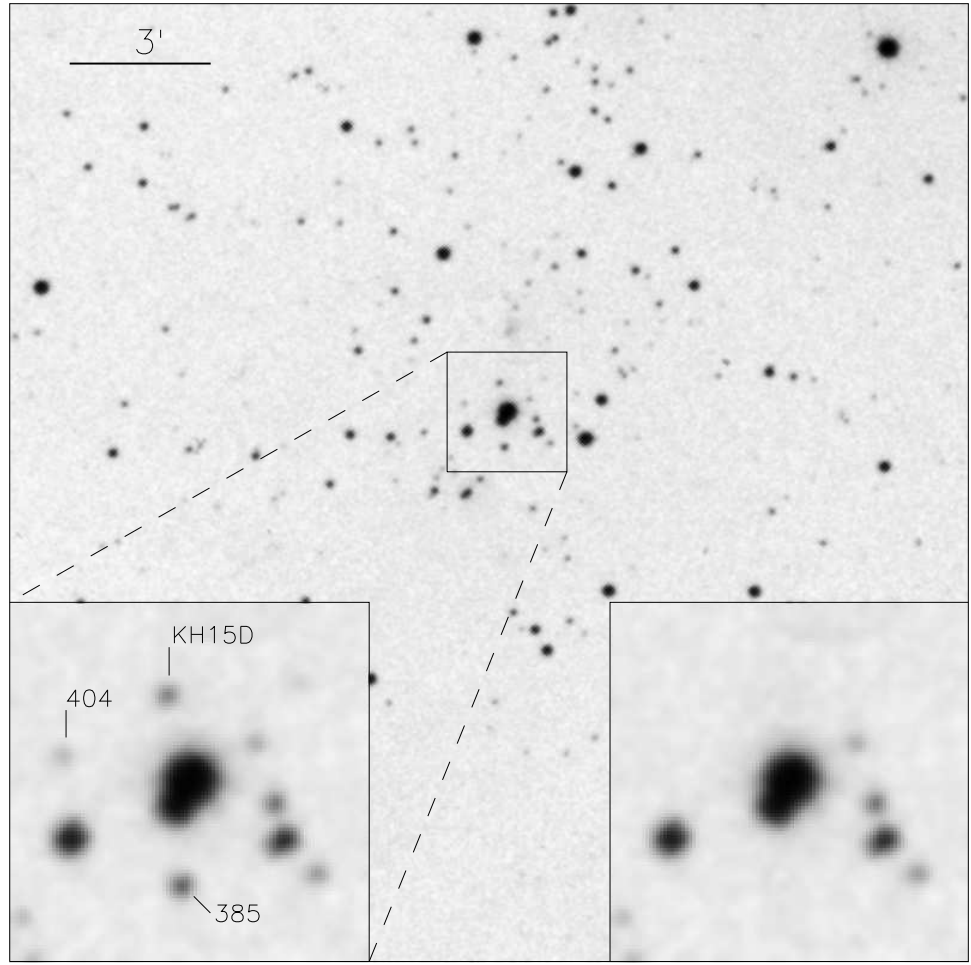}
\caption{Digitization of a I-N/RG5 plate from 1968 Jan 21. North
is up and east is left. The field of view is
20\arcmin$\times$20\arcmin, centered on the bright star HD~47887.
Insets in the corners are close-ups of the region surrounding HD~47887, 
showing KH~15D and two of the reference stars before and after
subtracting the PSF model (Eqn.~\ref{eq:pdp}). \label{field}}
\end{center}
\end{figure}

We digitized these plates at Asiago Observatory using a commercial
flatbed scanner with a resolution of 1600 dots per inch (dpi) and a
dynamic range of 14 bits~pixel$^{-1}$. The pixel scale of the
digitized images is 1.53 arcseconds~pixel$^{-1}$. Given the typical
seeing was 6--9\arcsec, the digitized stellar images have a full width
at half maximum of 4--6 pixels. The entire $5^{\circ} \times
5^{\circ}$ field of view was scanned, but for our analysis we
extracted a $20\arcmin \times 20\arcmin$~subraster centered on
HD~47887. A sample digitized image is shown in Figure \ref{field}.

\section{Photometry}
\label{photometry}

Astronomical photographic emulsions used throughout the last century
typically consisted of a suspension of silver halide in a substrate of
gelatin \citep{sw62}. When light strikes the halide, it produces a
distribution of metallic silver grains that is made permanent by
chemical development and fixation. The challenge of photographic
stellar photometry is to relate the spatial distribution of the grains
in a stellar image to the flux of the star.

The scanner measures $T(x,y)$, the fraction of transmitted light as a
function of position on the plate. What is actually recorded is the
``density distribution,''
\begin{equation}
d(x, y) = 2^{N_{\rm bit}}\left[1 - T(x, y)\right],
\end{equation}
where $N_{\rm bit}=14$ is the number of bits per pixel in the
digitized image. The relationship between $d(x,y)$ and stellar flux is
nonlinear. In particular, for intensities above some threshold value,
the density saturates. The following sections describe our method to
correct for the effects of nonlinearity and saturation, and to
calibrate the magnitude of KH~15D using a system of local reference
stars on each plate.

\subsection{Point-spread-function fitting}

Our first step was to fit a parametric model to the density
distribution of each star. The density distribution of faint stars is
well described by a two-dimensional Gaussian function,
\begin{equation}
d_G(x, y) = d_{\rm sky} +
            \frac{d_{\rm peak}}{2\pi \sigma_x \sigma_y \sqrt{1-\rho^2}}
            \exp\left(- \frac{q^2}{2(1-\rho^2)} \right),
\label{eq:gaussian}
\end{equation}

\begin{figure}[p]
\begin{center}
\includegraphics[width=1\textwidth]{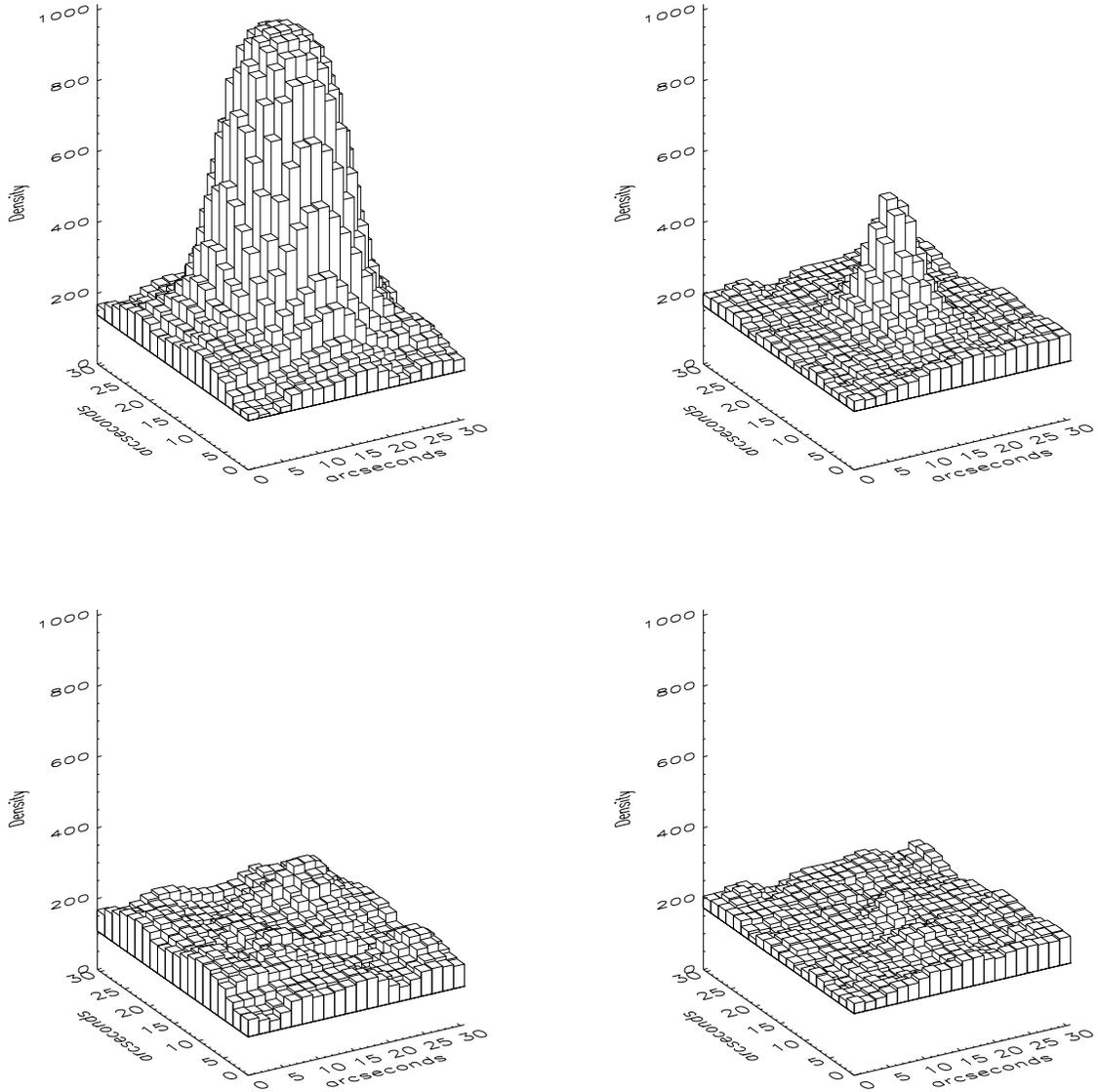}
\caption{Example density distributions from the plate shown in
Fig.~1. The upper panels show a bright star (left)
and KH~15D (right). The lower panels show the residuals after
subtracting the PSF model (Eqn.~\ref{eq:pdp}). \label{bright}}
\end{center}
\end{figure}

\noindent where $\rho$ parameterizes the rotation and the dimensionless
elliptical coordinate $q$ is defined via the equation
\begin{equation}
q^2 =
\frac{(x-x_0)^2}{\sigma_x^2} -
\frac{2\rho(x-x_0)(y-y_0)}{\sigma_x \sigma_y} +
\frac{(y-y_0)^2}{\sigma_y^2}.
\label{eq:q}
\end{equation}
In contrast, bright stars have a flat-topped density distribution, due
to saturation. The top two panels of Fig.~\ref{bright} show the
density distributions of a bright star ($I \sim 9$) and the much
fainter KH~15D, from the plate shown in Fig.~\ref{field}. In order to
accomodate a wide range of stellar magnitudes, one needs a function
that changes shape with brightness. We obtained good results with a
function presented by \citet{s79},
\begin{equation}
d_S(x,y) = \left[ d_G(x,y)^{-q} + d_{\rm sat}^{-q} \right]^{-1/q},
\label{eq:pdp}
\end{equation}
where $d_{\rm sat}$ is the saturation level of the plate, and $q$
describes the sharpness of the transition to saturation. This function
has the desirable limits $d_S \rightarrow d_G$ for $d_G \ll d_{\rm
sat}$, and $d_S \rightarrow d_{\rm sat}$ for $d_G \gg d_{\rm sat}$. It
is specified by 9 parameters: $x_0$ and $y_0$; $d_{\rm sky}$ and
$d_{\rm peak}$; $\sigma_x$, $\sigma_y$, and $\rho$; and the saturation
parameters $d_{\rm sat}$ and $q$.

The bottom two panels of Fig.~\ref{bright} show the residuals after
fitting Eqn.~\ref{eq:pdp} to the two example stars. In both cases the
fit is excellent. For the range of stellar brightnesses considered in
this paper, the peak residual was always $\lesssim$10\% of the peak
value of the density distribution. After fitting each star, we used
the best-fitting values of the parameters to compute
\begin{equation}
S \equiv d_{\rm peak} \times \sigma_x \sigma_y,
\label{eq:total_density}
\end{equation}
which is proportional to the volume of the density distribution above
the sky level, in the limit $d_{\rm sat}\rightarrow \infty$. By
computing $S$ in this manner, we attempt to correct for saturation and
provide a quantity that is easily related to total stellar flux.

We selected 163 preliminary reference stars from the
\citet[][hereafter F99]{fms+99} catalog of $BVRI$ CCD photometry of
NGC~2264.  Our selection criteria were that the stars should fall
within the $20\arcmin \times 20$\arcmin~square region centered on
HD~47887, lack close neighbor stars, and span a range of magnitudes
($12.5 < I < 16$) and colors ($0.5 < V-I < 3.0$) bracketing KH~15D
($I=14.4$, $V-I = 1.6$).

For each star on each plate, we performed a nonlinear least-squares
fit of Eqn.~\ref{eq:pdp} to the $20\times 20$~pixel
($\approx$30$\arcsec \times 30\arcsec$) region surrounding the
centroid of the density function.  At first, the ``plate parameters''
$q$, $D_s$, $\sigma_x$, $\sigma_y$, and $\rho$ were determined
independently for each star on a given plate. Then each plate
parameter was fixed at the median of the values obtained for all stars
on that plate, and the stellar fits were re-computed without varying
the plate parameters. Finally, we computed $S$ for each star.

In practice, we found that KH~15D and many reference stars were well
described by the simple Gaussian function
(Eqn.~\ref{eq:gaussian}). The results reported in this paper are based
on the analysis using Eqn.~\ref{eq:pdp}, but we also repeated the
analysis with the Gaussian function after dropping the brightest
reference stars. None of our conclusions changed significantly.

\subsection{Reduction to standard magnitudes}
\label{subsec:magnitudes}

\begin{figure}[p]
\begin{center}
\includegraphics[width=1\textwidth]{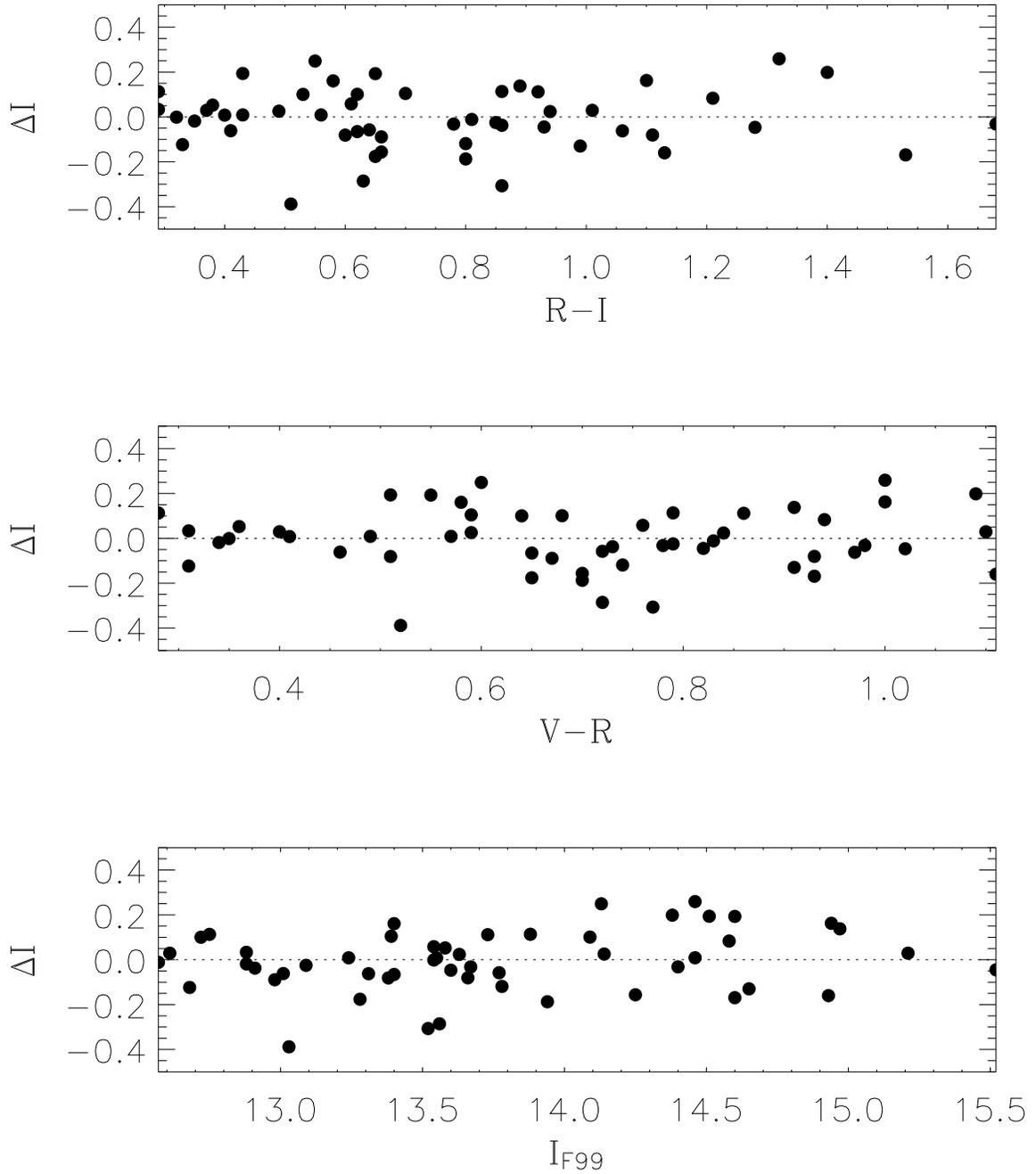}
\caption {The differences between fitted $I$ magnitudes and catalog
(F99) $I$ magnitudes, as a function of F99 magnitude and colors.
There are no significant correlations between the residuals and either
color or magnitude, which is why we did not include color terms or
higher-order nonlinear terms in
Eqn.~\ref{eq:mag_equation}. \label{resid_vs_color}}
\end{center}
\end{figure}

Next, we sought the relationship between $S$ and standard
magnitudes. For the I-N/RG5 plates, we converted $S$ into Cousins $I$
using the relation
\begin{equation}
I = I_0 - 2.5(1 + \epsilon) \log_{10}{S},
\label{eq:mag_equation}
\end{equation}
where $\epsilon$ represents the nonlinearity: the stellar flux is
proportional to $S^{1+\epsilon}$.  For each plate, we determined the
values of $I_0$ and $\epsilon$ that minimized the sum of squared
residuals between the fitted magnitudes and the F99 magnitudes. We
experimented with higher-order nonlinear terms, and with color terms,
but found that these were unhelpful. The residuals show no significant
correlation with either magnitude or color (see
Fig.~\ref{resid_vs_color}). For most of the plates, $\epsilon$ was
about 0.3.

Many of the reference stars are cluster members which, as young stars,
are likely to be variable. At this stage, we culled the most highly
variable stars from the list. We computed the standard deviation
$\sigma_I$ of each $I$ band time series, and rejected 110 stars with
$\sigma_I > 0.2$~mag. The remaining 53 stars were the final set of
reference stars used to determine the magnitude of KH~15D. The choice
of 0.2~mag is somewhat arbitrary; increasing the cutoff to 0.3~mag
doubles the number of reference stars but has no systematic effect on
the derived magnitudes of KH~15D.

\begin{figure}[!h]
\begin{center}
\includegraphics[width=.8\textwidth]{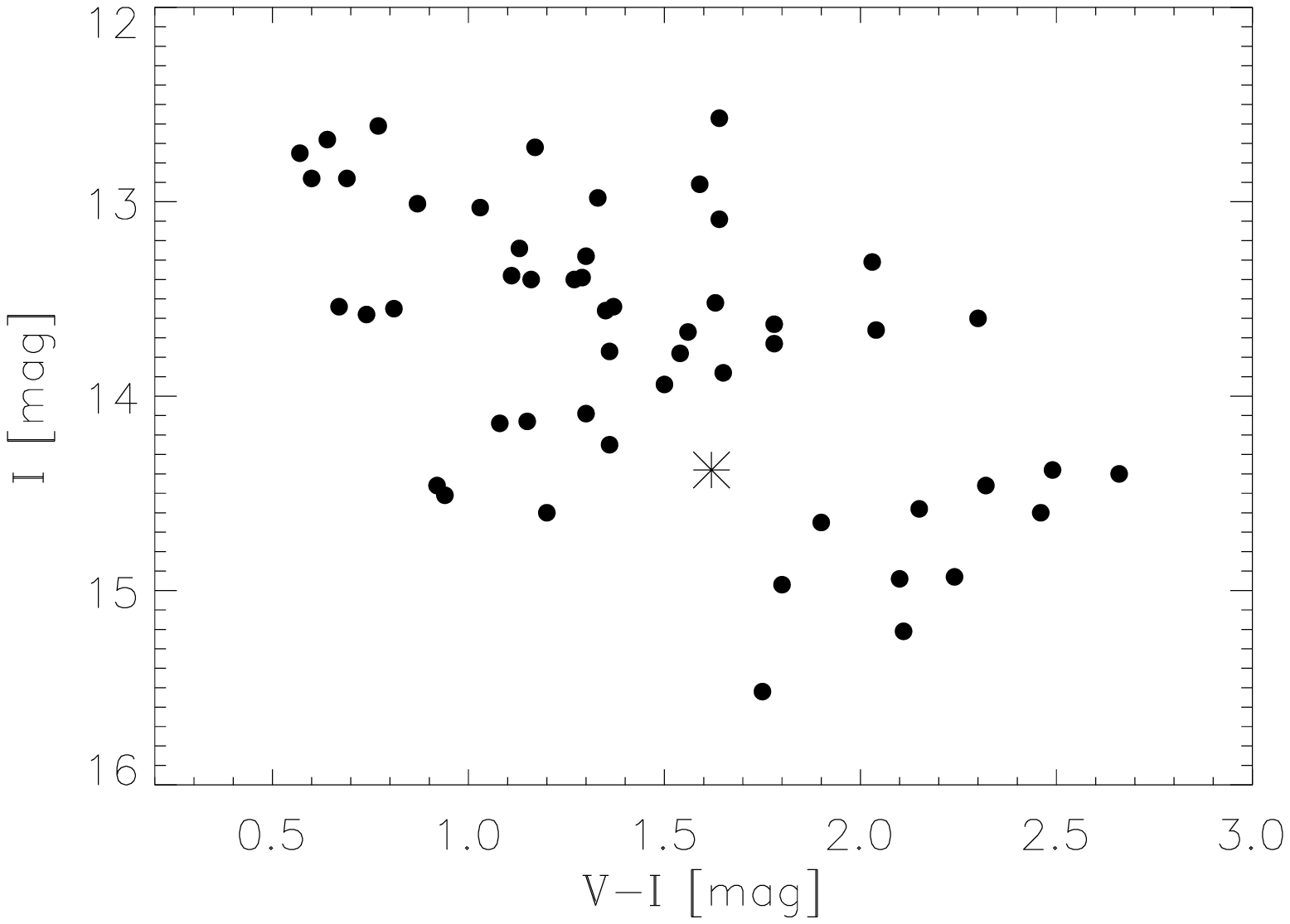}
\caption {A color--magnitude diagram for the 53 reference stars
(filled circles), and KH~15D (asterisk). Data taken from
F99. \label{colormag}}
\end{center}
\end{figure}

Figure \ref{colormag} is a color-magnitude diagram of the 53 reference
stars and KH~15D, showing that KH~15D is bracketed in both color and
magnitude. For each reference star, Table~\ref{stats_table} gives the
F99 catalog number, F99 $I$ magnitude, the mean $I$ magnitude measured
in the Asiago plates, the difference between those two magnitudes, and
$\sigma_I$.

For the three 103a-E/RG1 plates, we used the same set of 53 reference
stars to convert $S$ into Johnson $R$ band, using a relation analogous
to Eqn.~\ref{eq:mag_equation}. Likewise, for the single 103a-O/GG5
plate, we converted $S$ into Johnson $B$ band. The entries for these
plates in Table~\ref{stats_table} are estimates of $I$ under the
assumption that the colors of KH~15D were $R-I=0.80$ and $B-I=2.94$,
as measured by F99 (see also \S~\ref{sec:color}).

\subsection{Estimation of errors}
\label{error}

\begin{figure}[!h]
\begin{center}
\includegraphics[width=.8\textwidth]{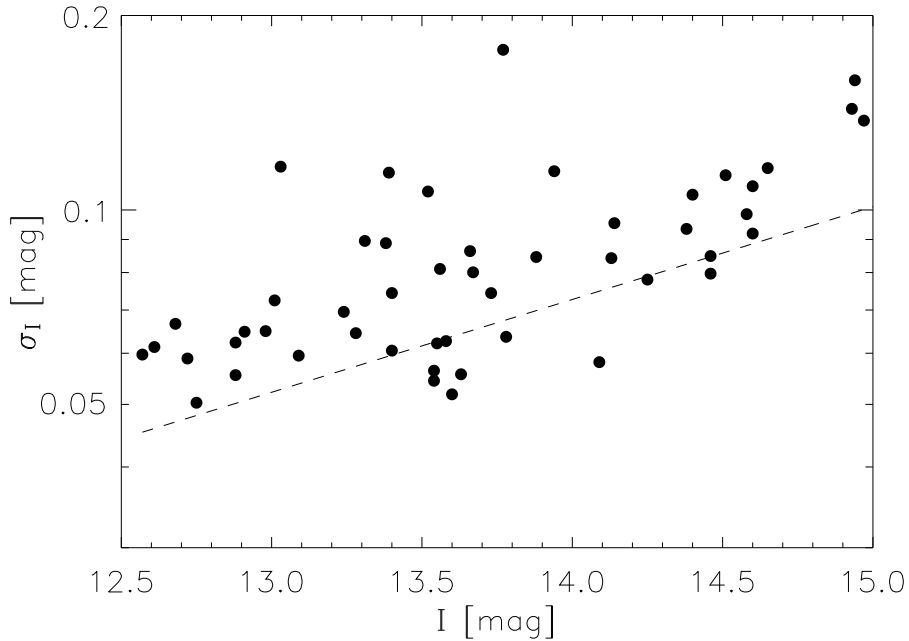}
\caption {RMS scatter in the time series of each standard star, as a
function of mean $I$ magnitude.  The dashed line is our approximation
$\sigma_{I, {\rm min}}(I)$ for the lower envelope. \label{errplot} } 
\end{center}
\end{figure}

Figure~\ref{errplot} is a plot of $\sigma_I$ vs.\ the time-averaged
$I$ magnitude for all the reference stars. We interpret the lower
envelope of the points in Figure~\ref{errplot} as the limiting
uncertainty of our relative $I$ band measurements, as a function of
magnitude. Reference stars for which $\sigma_I$ is much larger than
this envelope are presumably variable stars. A reasonable
approximation of the lower envelope is
\begin{equation}
\log{\sigma_{I, {\rm min}}}(I) = -0.990 + 0.144(I-14.4),
\end{equation}
which is plotted as a dashed line in Fig.~\ref{errplot}.

The uncertainty in a single measurement on a given plate may be
different than $\sigma_{I, {\rm min}}$, depending on the quality of
that plate. One measure of plate quality is $\sigma_m$, the standard
deviation of residuals to the fit of Eqn.~\ref{eq:mag_equation}, in
magnitudes. We supposed that the uncertainty is proportional to
$\sigma_m$. Thus, for a star with magnitude $I$ measured on a plate
with fit error $\sigma_m$, we estimated the uncertainty to be
\begin{equation}
{\rm Unc.} = \sigma_{I, {\rm min}}(I) \times 
             \left( \frac{\sigma_m} {\sigma_{m,{\rm min}}} \right),
\label{eq:errors}
\end{equation}
where $\sigma_{m,{\rm min}}$ is the minimum value of $\sigma_m$ among
all the plates. For the three $R$-band measurements and the single
$B$-band measurement, for which a long time series was not available,
we used $\sigma_m$ as the estimate of uncertainty. 

\section{Results}
\label{results}

\begin{figure}[!b]
\begin{center}
\includegraphics[width=.8\textwidth]{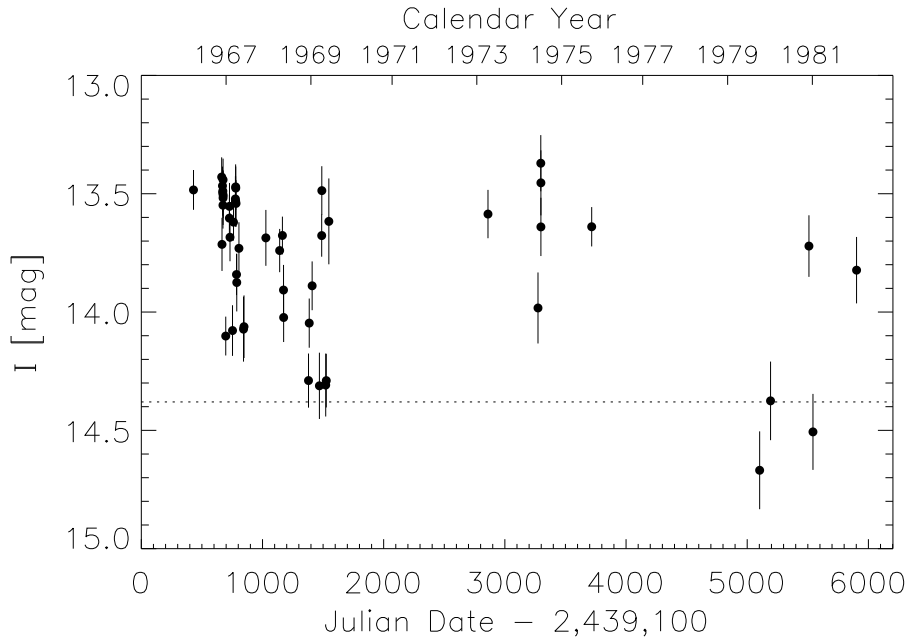}
\caption {$I$ band light curve of KH~15D from the Asiago plates. The
dotted line markes the magnitude of the bright state as observed
today. The error bars were computed with
Eqn.~\ref{eq:errors}. \label{comparo} }
\end{center}
\end{figure}

The resulting $I$ band magnitudes of KH~15D are given in
Table~\ref{kh_table}. Figure~\ref{comparo} shows the $I$ band light
curve, in which there are significant variations. In the following
sections we compare the Asiago light curve to the modern light curve.

\subsection{Periodicity}
\label{periodicity}

\begin{figure}[!b]
\begin{center}
\includegraphics[width=.8\textwidth]{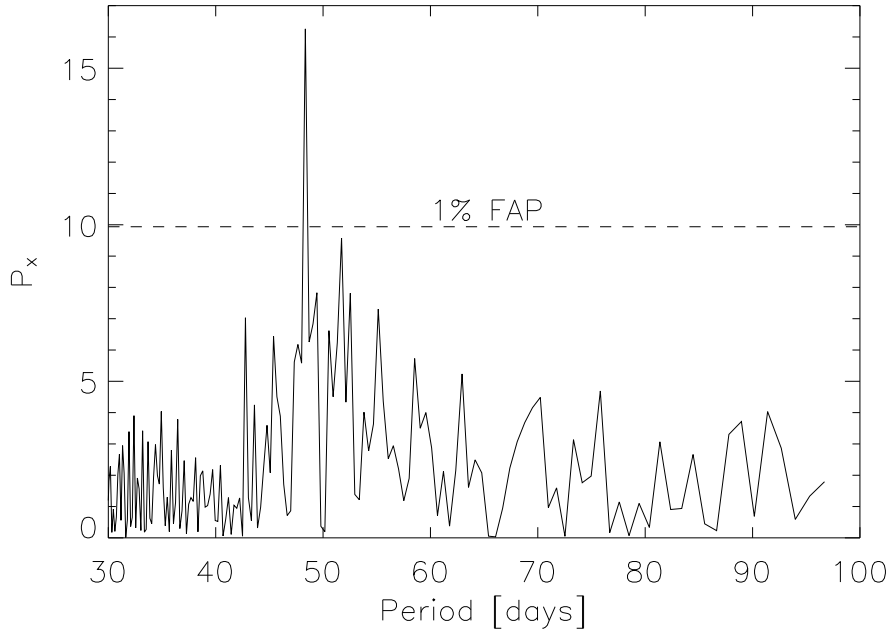}
\caption {Lomb-Scargle periodogram of the light curve shown in
Fig.~\ref{comparo}. Peaks above the dashed line have $<$1\% chance
of being a false alarm. \label{periodogram} }
\end{center}
\end{figure}

The brightness variations were periodic in the past, as they are
today. A Lomb-Scargle periodogram of the $I$ band light curve
(Fig.~\ref{periodogram}) shows a highly significant peak at a period
of 48.42 days, which is close to the modern period ($48.35\pm
0.02$~days; Herbst et al.\ 2002).  To estimate the statistical
uncertainty in our period determination, we used a Monte Carlo
procedure. Assuming the measurement errors are Gaussian with standard
deviations given by the recipe of \S~\ref{error}, we generated $10^4$
statistical realizations of the $I$ band light curve and determined
the peak value of the periodogram in each case. The resulting
distribution of periods had a standard deviation of 0.02 days. Thus,
both our period and the modern period have a formal uncertainty of
$\sigma=0.02$~days and differ by $3.5\sigma$, or $1.75\sigma$ in each
period. This may indicate a discrepancy, but it is also quite possible
that our Monte Carlo procedure underestimates the true uncertainty,
due to the time evolution of the light curve.

\begin{figure}[!b]
\begin{center}
\includegraphics[width=1\textwidth]{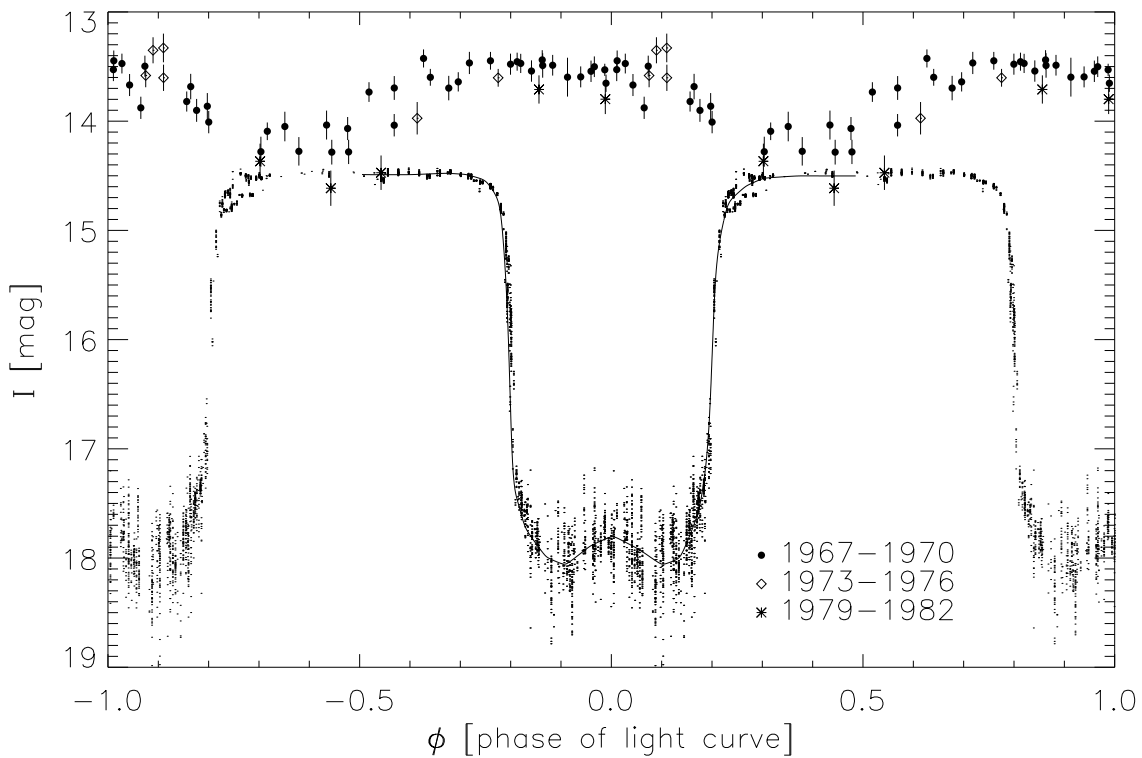}
\caption { Phased light curve of KH~15D from 1967-1982 (filled
circles) and 2001-2002 (small dots).  The solid line is a B-spline
interpolation of the 2001-2002 data (see
\S~\ref{discussion}).  \label{lc_combo} }
\end{center}
\end{figure}

We computed the phase of each observation,
\begin{equation}
\phi = ({\rm J.D.} - J_0) \mbox{ mod } P,
\label{phase_eqn}
\end{equation}
where J.D.\ is the Julian date of the observation, $P=48.35$~days, and
$J_0=2,452,690.67$, taken from the most recent ephemeris (Hamilton
2004, in preparation). Figure~\ref{lc_combo} shows the phased light
curve. For comparison, we have also plotted CCD-based measurements
from 2001--2002, kindly provided by C.\ Hamilton.

\subsection{Brightness variations}

In the Asiago light curve, the star alternates periodically between a
bright state and a faint state. The fading events last $\approx$20
days, which is about the same duration as the 2000--2001 eclipses.
However, the fractional variation in flux is much smaller in the
Asiago light curve. The difference between the average magnitude in
the bright state (defined as $-0.30 \le \phi \le 0.10$) and the faint
state ($0.25 \le \phi \le 0.55$) is $\Delta I = 0.67\pm 0.07$~mag.  
By contrast, the modern eclipse depth is
3.5~magnitudes. Interestingly, the three faintest measurements are
also from the most recent time 
series (1979--1982; plotted with asterisks in Fig.~\ref{lc}), which is
consistent with a progessive deepening of the eclipses.

\subsection{Magnitude of the bright state}
\label{sec:bright_state}

The mean magnitude of the bright state in the Asiago light curve is
$I=13.57\pm 0.03$, as compared to $14.47\pm 0.03$ for the modern
data. The bright state was formerly 0.90 magnitude (2.3 times)
brighter than it is today. One might reasonably wonder whether this
surprising discrepancy is due to an error in the zero point of our
magnitude relation, or a systematic effect such as contamination by
scattered light from HD~47887.

\begin{figure}[!t]
\begin{center}
\includegraphics[width=.8\textwidth]{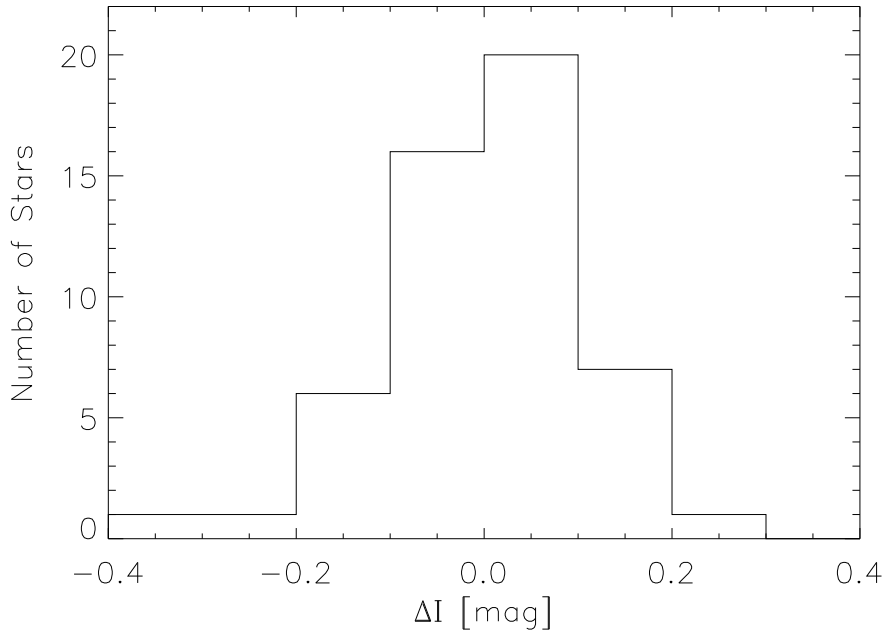}
\caption {Histogram of differences between mean $I$ magnitudes from
the Asiago plates, and catalog $I$ magnitudes (F99), for the 53
reference stars. The distribution is centered near zero and has
standard deviation 0.14. \label{deltam}}
\end{center}
\end{figure}

Regarding the accuracy of the zero point, Fig.~\ref{deltam} shows a
histogram of $\Delta I$, the differences between the time-averaged
magnitudes of the 53 reference stars in the Asiago plates, and the
corresponding F99 magnitudes. More than 67\% of the stars have $\Delta
I<0.1$~mag, and the maximum $\Delta I$ is 0.3~mag. This gives us
confidence that the zero point is accurate to $\approx$0.15~mag.

Regarding possible contamination, the I-N/RG5 plates were chosen
precisely to minimize this problem. In the top right panel of Fig.~2,
the halo of HD~47887 is evident as a small gradient in the sky level
surrounding KH~15D. This gradient is too small to cause a 0.9~mag
(230\%) error in the estimate of the volume beneath the peak. Star 385
is even closer to HD~47887 (see Fig.~1) and has $\Delta I=
-0.05$~mag. Star 404 is also within the halo of scattered light, and
has $\Delta I=0.18$~mag, despite being a known variable star
\citep{kkp+71}.

\subsection{Color variation}
\label{sec:color}

It would be interesting to know whether the decrease in overall
brightness was accompanied by a color change. We have very limited
color information, having concentrated on the I-N/RG5 plates, but the
few bluer plates in our sample show no evidence for a color change. In
particular, we obtained estimates of $B$ and $R$ from plates exposed
less than one hour apart on 1974~December~15 (J.D.\ 2,439,774.6; see
Table \ref{kh_table}). The phase on this date was $\phi = 0.11$,
corresponding to the transition from the bright state to the faint
state. The result was $B-R=2.0\pm 0.3$, in agreement with the F99
measurement, $B-R=2.14$.

\subsection{Phase of minimum light}

In the 30 years between the earliest group of Asiago observations and
the determination of the \citet{hhv+02} ephemeris, there have been more
than 200 eclipses. The 0.02-day uncertainty in each period causes a
phase uncertainty of $\Delta\phi \approx 0.1$ in the connection
between the modern light curve and the Asiago light curve. Yet the
phase of minimum light in the Asiago light curve occurs at
$\phi_m=0.4$, rather than $\phi_m = 0$ as in the modern light
curve. This implies, at the 4$\sigma$ level, that there has been a
shift in the phase of minimum light.

An important caveat, as mentioned in \S~\ref{periodicity}, is that
period determination for this system is difficult because the light
curve does not repeat exactly from period to period. We believe it is
reasonable that the period uncertainty has been underestimated,
despite our best effort and the best effort of Herbst et al.\ 2002.
This makes us reluctant to attach too much significance to the
apparent phase shift before this point is clarified with additional
archival data or continued monitoring.

\section{Summary and discussion}
\label{discussion}

The fading events of KH~15D were occurring between 1967 and 1982 with
nearly the same period and duration as observed today. However, the
system has changed over the past 30 years in two major
respects. First, the bright state has decreased in flux by a factor of
2.3 ($0.9\pm 0.1$~mag). Second, the contrast between the bright state
and the faint state has increased dramatically. Formerly, the flux of
the faint state was 54\% of the flux of the bright state ($\Delta
I=0.67\pm 0.07$~mag). In modern observations, the corresponding figure
is 4\% (3.5~mag).

Interestingly, these two observations could both be explained by a
time-independent flux that was present during the Asiago observations
and not present during the modern observations. A steady light source
with 1.3 times the flux of the eclipsing K7 star would increase the
total flux by a factor of 2.3, and would also dilute the
eclipses. When the K7 star is totally eclipsed, the total flux would
be reduced to $1.3/2.3 = 57\%$ of the bright state.

\begin{figure}[p]
\begin{center}
\includegraphics[width=1\textwidth]{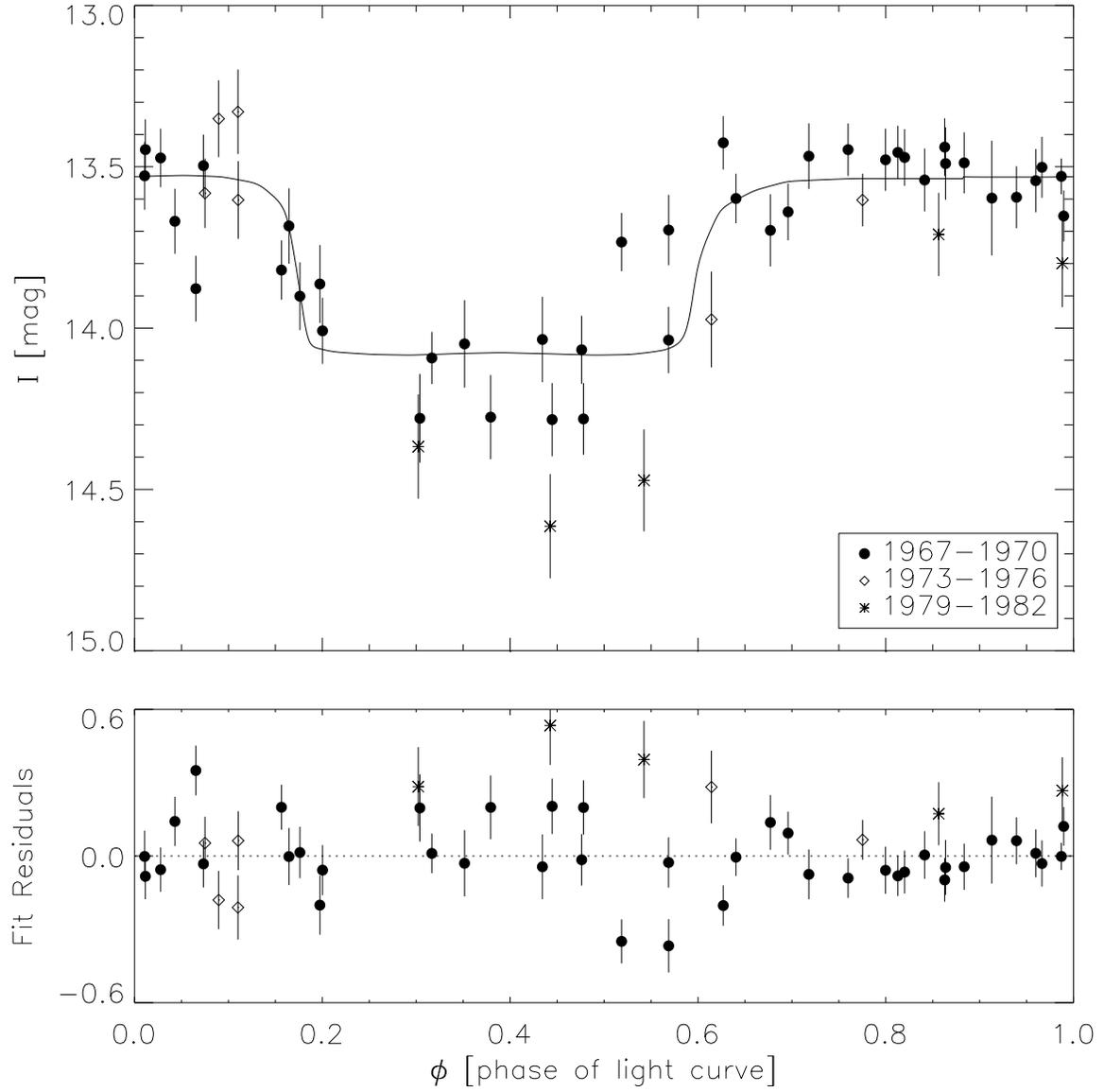}
\caption {Phased light curve of KH~15D from 1967--1982, with a model
based on a transformation of the 2000--2001 phased light curve (see
\S~\ref{discussion}). The residuals (model$-$data) are plotted beneath
the light curve. \label{lc}}
\end{center}
\end{figure}

To elaborate upon this idea, we created a smooth model of the
2000--2001 light curve using B-spline interpolation. This model is
shown as a solid line in Figure~\ref{lc_combo}. Then we considered
three possible transformations of the model light curve: (1) add a
constant light source of magnitude $I_0$; (2) shift the phase by
$\Delta \phi$; and (3) stretch or compress the eclipse duration by a
factor $R$. We determined the values of $I_0$, $\Delta\phi$, and $R$
that provide the best match to the Asiago light curve, using a
nonlinear least-squares algorithm. The best fit was obtained for $I_0
= 14.11$, $\Delta \phi = 0.38$, and $R = 0.9$. In Figure~\ref{lc}, the
transformed modern light curve is superposed on the Asiago data, and
the residuals are plotted beneath the data. The RMS scatter of the
residuals is 0.17~mag and the reduced $\chi^2$ is 2.1.

The success of this simple model leads us to hypothesize that there is
a second star in the KH~15D system. The second star was formerly
blended with the $I=14.4$ K7 star seen today, but is now completely
obscured. This provides an appealing explanation of the Asiago light
curve but does raise an obvious question: where did the second star
go?

The most economical answer, in the sense that it requires the fewest
new complexities, is that the second star is currently behind the same
opaque material that causes the periodic eclipses of the K7 star. The
composition and arrangement of that material are unknown. Some
theories are a nearly edge-on circumstellar or protoplanetary disk
\citep{hhs+01, hhv+02, wgs+03, abw+03}, a dusty banana-shaped vortex
\citep{bv03}, and a cometary distribution of accreting material around
a low-mass companion \citep{gt02}. The material might be distributed
in such a way that one star is always hidden and the other star comes
into view periodically, due to the orbital motion of the KH~15D or the
circumstellar material.

If the system is truly a binary, one would expect to see radial
velocity variations. Assuming the observed K7 star has a mass of
$\approx$0.6~$M_{\odot}$, as estimated by Hamilton et al.\ 2001, and
that it is in a 48-day edge-on orbit with a star of similar mass, its
orbital velocity would be $\approx$30~km~s$^{-1}$. This is ten times
larger than the velocity shift measured by \citet{hhm+03} between two
widely separated phases. Of course, it is possible that the full
radial velocity curve shows larger variations. It is also possible
that the second star is a wide binary companion, an unrelated
background or foreground star, or a binary companion in a nearly
face-on orbit.

Allowing the possibility of a second star to the system does not, by
itself, explain the eclipses or their evolution. However, our
discovery that KH~15D was brighter in the past puts two of the
previous observations of this system in a different light. One of
these is the result of the 1913--1951 Harvard plate
analysis. \citet{wgs+03} derived limits on the fraction of time that
the eclipses were $>$1~mag, but this ignored the possibility that the
system was brighter at {\it all} phases in the past. The more precise
statement of the result is that the system was rarely more than 1~mag
fainter than the {\it modern} bright state.  More speculatively, there
may be a connection to the surprising observation by \citet{kh98} that
during a few of the re-brightening events, the system became {\it
brighter} than the usual bright state by 0.1--0.5~mag. Perhaps during
these events we were allowed a peek at the second star.

With the Asiago plates we have discovered several new clues regarding
the mysterious eclipses of KH~15D, demonstrating the value of
long-term preservation of astronomical images. It will be interesting
to extend the historical analysis with additional plates from the
1980s and 1990s, to reveal when and how the extra light source turned
off, and to complete the connection with modern data.

\acknowledgements We are indebted to Francesca Rampazzi and Cesare
Barbieri for their kind assistance with the Asiago archive and their
hospitality during J.A.J.'s visit.  We are also grateful to Milcho
Tsvetkov for creating the Wide-Field Plate
Database\footnote{http://www.skyarchive.org}, which made it easy to
identify promising plate collections. Geoff Marcy, Kris Stanek, and
Dimitar Sasselov provided encouragement and comments on the manuscript
which were much appreciated. J.N.W.\ is supported by an NSF Astronomy
\& Astrophysics Postdoctoral Fellowship under grant AST-010347.

\clearpage

\clearpage

\begin{deluxetable}{ccccc}
\tablecaption{Photometric reference stars in NGC~2264\label{stats_table}}
\tablewidth{0pt}
\tablehead{
\colhead{F99} &
\colhead{$I_{\rm F99}$} &
\colhead{$I_{\rm Asiago}$\tablenotemark{a}} & 
\colhead{$\Delta I$\tablenotemark{b}} &
\colhead{$\sigma_{I}$} \\
\colhead{Catalog No.} &
\colhead{} &
\colhead{} &
\colhead{} & 
\colhead{}
}
\startdata
  113 & 15.52 & 15.51 &  0.01 & 0.24\\
  122 & 14.51 & 14.44 &  0.07 & 0.11\\
  128 & 14.60 & 14.48 &  0.12 & 0.09\\
  160 & 14.13 & 13.85 &  0.28 & 0.08\\
  165 & 13.54 & 13.50 &  0.04 & 0.06\\
  173 & 14.65 & 14.59 &  0.06 & 0.12\\
  195 & 13.38 & 13.48 & -0.10 & 0.09\\
  208 & 13.73 & 13.70 &  0.03 & 0.07\\
  227 & 13.39 & 13.30 &  0.09 & 0.11\\
  263 & 13.40 & 13.29 &  0.11 & 0.06\\
  272 & 12.57 & 12.57 &  0.00 & 0.06\\
  281 & 13.60 & 13.58 &  0.02 & 0.05\\
  289 & 14.94 & 14.95 & -0.01 & 0.16\\
  297 & 14.40 & 14.44 & -0.04 & 0.11\\
  305 & 13.31 & 13.32 & -0.01 & 0.09\\
  315 & 13.28 & 13.43 & -0.15 & 0.06\\
  320 & 12.88 & 12.94 & -0.06 & 0.06\\
  321 & 13.03 & 13.33 & -0.30 & 0.12\\
  338 & 13.58 & 13.57 &  0.01 & 0.06\\
  342 & 13.94 & 14.10 & -0.16 & 0.11\\
  346 & 14.60 & 14.64 & -0.04 & 0.11\\
  353 & 13.63 & 13.63 &  0.00 & 0.06\\
  364 & 15.21 & 15.27 & -0.06 & 0.12\\
  374 & 13.52 & 13.75 & -0.23 & 0.11\\
  378 & 14.25 & 14.33 & -0.08 & 0.08\\
  381 & 13.09 & 13.12 & -0.03 & 0.06\\
  385 & 12.91 & 12.96 & -0.05 & 0.06\\
  404 & 14.97 & 14.79 &  0.18 & 0.14\\
  409 & 13.88 & 13.74 &  0.14 & 0.08\\
  411 & 14.38 & 14.26 &  0.12 & 0.09\\
  419 & 13.77 & 13.91 & -0.14 & 0.18\\
  422 & 13.66 & 13.67 & -0.01 & 0.09\\
  424 & 13.78 & 13.79 & -0.01 & 0.06\\
  426 & 13.56 & 13.72 & -0.16 & 0.08\\
  430 & 14.93 & 14.97 & -0.04 & 0.14\\
  431 & 13.01 & 12.97 &  0.04 & 0.07\\
  432 & 14.46 & 14.38 &  0.08 & 0.08\\
  434 & 12.98 & 13.07 & -0.09 & 0.06\\
  440 & 13.67 & 13.63 &  0.04 & 0.08\\
  443 & 14.46 & 14.31 &  0.15 & 0.08\\
  444 & 12.72 & 12.70 &  0.02 & 0.06\\
  450 & 14.09 & 14.02 &  0.07 & 0.06\\
  451 & 12.75 & 12.70 &  0.05 & 0.05\\
  452 & 14.58 & 14.45 &  0.13 & 0.10\\
  453 & 12.61 & 12.65 & -0.04 & 0.06\\
  460 & 13.24 & 13.22 &  0.02 & 0.07\\
  462 & 13.54 & 13.48 &  0.06 & 0.05\\
  463 & 12.88 & 12.79 &  0.09 & 0.06\\
  481 & 13.40 & 13.45 & -0.05 & 0.07\\
  504 & 13.55 & 13.53 &  0.02 & 0.06\\
  512 & 14.14 & 14.15 & -0.01 & 0.10\\
  515 & 12.68 & 12.78 & -0.10 & 0.07\\
\tablenotetext{a}{Time average from all plates.}
\tablenotetext{b}{$\Delta I = I_{\rm F99} - I_{\rm Asiago}$}
\enddata
\end{deluxetable}

\clearpage

\begin{deluxetable}{ccc}
\tablecaption{Photometric Measurements of KH~15D \label{kh_table}}
\tablewidth{0pt}
\tablehead{
\colhead{J.D.} &
\colhead{$\phi$~\tablenotemark{a}} &
\colhead{I}
}
\startdata
2439530.46 & 0.8121 & $13.46 \pm 0.08$\\
2439763.22 & 0.6261 & $13.43 \pm 0.08$\\
2439765.65 & 0.6762 & $13.7 \pm 0.1$\\
2439769.65 & 0.7591 & $13.45 \pm 0.08$\\
2439771.58 & 0.7989 & $13.5 \pm 0.1$\\
2439772.56 & 0.8192 & $13.47 \pm 0.09$\\
2439773.59 & 0.8405 & $13.5 \pm 0.1$\\
2439774.62 & 0.8619 & $13.44 \pm 0.09$\\
2439774.67 & 0.8628 & $13.5 \pm 0.1$\\
2439775.63 & 0.8826 & $13.49 \pm 0.09$\\
2439796.57 & 0.3157 & $14.09 \pm 0.08$\\
2439826.67 & 0.9384 & $13.6 \pm 0.1$\\
2439827.66 & 0.9588 & $13.5 \pm 0.1$\\
2439831.69 & 0.04213 & $13.7 \pm 0.1$\\
2439852.63 & 0.4753 & $14.1 \pm 0.1$\\
2439860.58 & 0.6396 & $13.60 \pm 0.08$\\
2439876.33 & 0.9655 & $13.50 \pm 0.09$\\
2439877.33 & 0.9862 & $13.53 \pm 0.06$\\
2439878.48 & 0.009931 & $13.5 \pm 0.1$\\
2439878.52 & 0.01074 & $13.45 \pm 0.09$\\
2439879.30 & 0.02694 & $13.47 \pm 0.09$\\
2439881.51 & 0.07251 & $13.5 \pm 0.1$\\
2439885.53 & 0.1556 & $13.82 \pm 0.09$\\
2439887.50 & 0.1965 & $13.9 \pm 0.1$\\
2439905.46 & 0.5680 & $13.7 \pm 0.1$\\
2439943.30 & 0.3506 & $14.0 \pm 0.1$\\
2439947.31 & 0.4335 & $14.0 \pm 0.1$\\
2440127.66 & 0.1636 & $13.7 \pm 0.1$\\
2440241.49 & 0.5178 & $13.73 \pm 0.09$\\
2440264.25 & 0.9886 & $13.65 \pm 0.08$\\
2440273.28 & 0.1753 & $13.9 \pm 0.1$\\
2440274.44 & 0.1994 & $14.0 \pm 0.1$\\
2440479.66 & 0.4439 & $14.3 \pm 0.1$\\
2440485.66 & 0.5679 & $14.0 \pm 0.1$\\
2440509.67 & 0.06442 & $13.9 \pm 0.1$\\
2440569.55 & 0.3030 & $14.3 \pm 0.1$\\
2440588.51 & 0.6951 & $13.64 \pm 0.09$\\
2440589.58 & 0.7172 & $13.5 \pm 0.1$\\
2440621.55 & 0.3785 & $14.3 \pm 0.1$\\
2440626.32 & 0.4772 & $14.3 \pm 0.1$\\
2440647.35 & 0.9121 & $13.6 \pm 0.2$\\
2441960.64 & 0.07422 & $13.6 \pm 0.1$\\
2442373.51 & 0.6135 & $14.0 \pm 0.2$\tablenotemark{b}\\
2442396.49 & 0.08877 & $13.4 \pm 0.2$\tablenotemark{b}\\
2442397.49 & 0.1093 & $13.3 \pm 0.2$\tablenotemark{c}\\
2442397.50 & 0.1095 & $13.6 \pm 0.2$\tablenotemark{b}\\
2442816.44 & 0.7744 & $13.60 \pm 0.08$\\
2444202.51 & 0.4418 & $14.6 \pm 0.2$\\
2444292.42 & 0.3013 & $14.4 \pm 0.2$\\
2444609.31 & 0.8555 & $13.7 \pm 0.1$\\
2444642.49 & 0.5417 & $14.5 \pm 0.2$\\
2445002.48 & 0.9872 & $13.8 \pm 0.1$\\
\enddata
\tablenotetext{a}{Phase calculated assuming a 48.35 day period.}
\tablenotetext{b}{103a-E/RG1 plate. $I$ was estimated assuming $R-I = 0.80$. }
\tablenotetext{c}{103a-O/GG5 plate. $I$ was estimated assuming $B-I = 2.92$. }
\end{deluxetable}

\end{document}